# Kerr-microresonator solitons from a chirped background

**Daniel C. Cole,**[1,2*] **Jordan R. Stone,**[1,2] **Miro Erkintalo,**[3] **Ki Youl Yang,**[4] **Xu Yi,**[4] **Kerry J. Vahala,**[4] **and Scott B. Papp**[1,2]

[1]*National Institute of Standards and Technology (NIST), Boulder, CO 80305, USA*
[2]*Department of Physics, University of Colorado, Boulder, CO 80309, USA*
[3]*Dodd-Walls Centre for Photonic and Quantum Technologies, Department of Physics, University of Auckland, Auckland 1142, New Zealand*
[4]*T. J. Watson Laboratory of Applied Physics, California Institute of Technology, Pasadena, CA 91125, USA*
*\*Corresponding author: daniel.cole@nist.gov*

**We demonstrate protected single-soliton formation and operation in a Kerr microresonator using a phase-modulated pump laser. Phase modulation gives rise to spatially varying effective loss and detuning parameters, which in turn lead to an operation regime in which multi-soliton degeneracy is lifted and a single soliton is the only observable behavior. Direct excitation of single solitons is indicated by observed reversal of the characteristic 'soliton step.' Phase modulation also enables precise control of the soliton pulse train's properties, and measured dynamics agree closely with simulations. We show that the technique can be extended to high repetition-frequency Kerr solitons through subharmonic phase modulation. These results facilitate straightforward generation and control of Kerr-soliton microcombs for integrated photonics systems.**

Dissipative temporal cavity solitons in Kerr microresonators [1–3] have the potential to provide the revolutionary capabilities of frequency combs in a chip-integrable platform. This would extend the reach of frequency combs to applications in communications, computation, and sensing with low size, weight, and power. Progress has come rapidly in the field of microresonator-soliton-based frequency combs, but for these combs to reach applications, simple, repeatable, and platform-independent methods of soliton generation and control are needed. The basic challenge is that solitons in microresonators are independent excitations, and a resonator can host zero, one, or many co-circulating solitons at a given pump-laser power and frequency, with each soliton giving rise to its own out-coupled pulse train. Further, under normal conditions microresonator solitons can only be generated by condensation from extended modulation-instability (MI) patterns (primary comb/Turing patterns, or noisy comb/spatiotemporal chaos) that provide appropriate initial perturbations. Thermal stability must be maintained as the intracavity power drops during the transition from a high duty-cycle MI pattern to a low duty-cycle soliton. A variety of schemes have been demonstrated to address these challenges and obtain single solitons [4–8], and many achieve excellent performance. In general these schemes increase experimental complexity, exploiting non-adiabatic variations in pump-laser power and frequency, and involve at least some amount of stochastic fluctuation in the output.

One notable possibility is modulation of the pump laser at a frequency near the resonator free-spectral range (FSR) [9–12], which can enable deterministic condensation of either one or zero solitons from an MI pattern. Further, it has been demonstrated that phase modulation (PM) can facilitate generation and control of single solitons [11,13,14]. Here we use PM at the FSR to deterministically excite single solitons directly from a chirped background, which remains otherwise stable, as proposed in Ref. [15]. Exiting the resonator is a train of solitons spaced by the round-trip time, as shown in Fig. 1a. Importantly, this scheme requires no transient perturbation to the system parameters.

Our results demonstrate a regime in which single-soliton operation is fundamentally protected, without the degeneracy between $N = 0, 1$, and many solitons that exists for a continuous-wave (CW) pump laser. To motivate the experimental work that follows, we present theoretical results that illustrate the utility of a PM pump. We use the nonlinear partial-differential Lugiato-Lefever equation (LLE) with modification of the driving term for phase modulation with depth $\delta_{PM}$ [1,15–18]:

$$\frac{\partial \psi}{\partial \tau} = -(1 + i\alpha)\psi + i|\psi|^2\psi - i\frac{\beta_2}{2}\frac{\partial^2 \psi}{\partial \theta^2} + Fe^{i\delta_{PM}\cos\theta}. \quad (1)$$

The normalized quantities used in the LLE are defined as follows [16]: $\psi$ is the envelope for the intracavity field normalized so that $|\psi|^2 = 1$ at the absolute threshold for parametric oscillation; $\tau = t/2\tau_{ph}$, where t is the time and $\tau_{ph} = 1/2\pi\Delta\nu$ is the cavity photon lifetime and $\Delta\nu$ is the cavity resonance linewidth; $\alpha = 2(\nu_0 - \nu_{pump})/\Delta\nu$ is the detuning between the pumped resonance with frequency $\nu_0$ and the pump laser with frequency $\nu_{pump}$; $F$ is the pump strength normalized so that $F^2 = 1$ at the absolute threshold for parametric oscillation; and $\beta_2 = -2D_2/2\pi\Delta\nu < 0$ is the anomalous resonator dispersion, with $D_2/2\pi = \partial^2\nu_\mu/\partial\mu^2|_{\mu=0} > 0$, where $\nu_\mu$ represents the set of cavity resonance frequencies. The azimuthal angle $\theta$ co-rotates at the frequency $f_{PM}$, which is presently assumed to be equal to the FSR.

We perform simulations of the LLE to investigate soliton degeneracy for the range of pump-laser detunings over which solitons exist. We use a fourth-order Runge-Kutta algorithm in the interaction picture [19] with adaptive step size [20]. The resulting soliton energy-level diagrams for the CW case ($\delta_{PM} = 0$) and the PM case ($\delta_{PM} = \pi$) are shown in Fig. 1b. We find that PM transforms the resonator excitation spectrum from a series of $N = 0, 1, 2, \ldots, N_{max}$ solitons to a single level $N = 1$ near threshold, eliminating soliton degeneracy. This occurs due to spatial variations of effective loss and detuning parameters that result from the phase modulation. We approximate these parameters by inserting the ansatz $\psi(\theta, \tau) = \phi(\theta, \tau)e^{i\delta_{PM}\cos\theta}$ into Eq. (1) [14]. By expanding the second-derivative term and setting derivatives of $\phi$ to zero we arrive at an equation for the quasi-CW background in the PM-pumped resonator:

$$F = (\gamma(\theta) + i\alpha_{eff}(\theta))\phi - i|\phi|^2\phi. \quad (2)$$

The effective loss and detuning terms are:

$$\gamma(\theta) = 1 + \frac{\beta_2}{2}\delta_{PM}\cos\theta, \quad (3)$$

$$\alpha_{eff}(\theta) = \alpha - \frac{\beta_2}{2}\delta_{PM}^2\sin^2\theta. \quad (4)$$

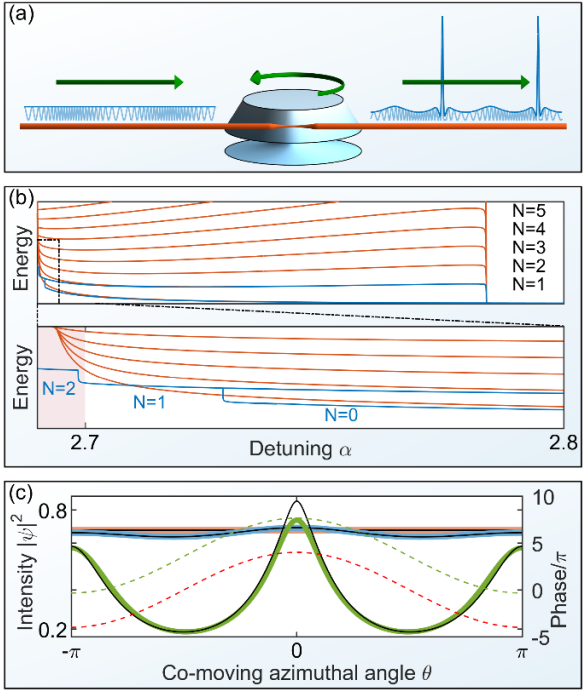

Fig 1. (a) Schematic for soliton generation in a PM-pumped resonator, neglecting interference in the output. (b) Simulated energy-level diagrams for the CW- (orange) and PM-pumped (blue, $\delta_{PM} = \pi$) resonator for $F^2 = 4, \beta = -0.0187$. With PM, an interval in $\alpha$ exists for which the single soliton is the only available energy level. This interval is fairly narrow, but we find that it is readily accessible in experiment. We also observe non-stationary states for values of $\alpha \leq 2.7$ in the PM case (red shading). (c) Simulated quasi-CW background intensity without (orange) and with PM of depth $\pi/2$ (blue) and $4\pi$ (green), with analytical approximations in black. Here $\alpha$ is slightly larger than the critical value for soliton formation. Dashed traces show the simulated phase profile of the field $\psi$ (green) and of the driving term $Fe^{i\delta_{PM}\cos\theta}$ (red) with modulation depth $4\pi$. The phase of the field is very nearly the phase of the drive plus a constant offset.

The field $\psi$ can be approximated using these local parameters:

$$\psi = \frac{Fe^{i\delta_{PM}\cos\theta}}{\gamma(\theta) + i\left(\alpha_{eff}(\theta) - \rho(\theta)\right)}, \quad (5)$$

where $\rho(\theta) = |\phi(\theta)|^2$ is the (smallest real) solution to the cubic polynomial in $\rho$ obtained from the modulus-square of Eq. (2). In neglecting the spatial derivatives of $\phi$ but retaining the derivatives of the term $e^{i\delta_{PM}\cos\theta}$ we make the approximation that the dominant effect of dispersion is its action on the existing phase-modulation spectrum. We note that a full analysis of the rich behavior of Eq. (1) without this approximation remains a promising avenue for future research.

Fig. 1c shows the predictions of simulations (color) and the analytical model (black). The two agree quantitatively for weak modulation ($\delta_{PM} = \frac{\pi}{2}$, blue) and qualitatively with larger depth ($\delta_{PM} = 4\pi$, green); both indicate that the field $\psi$ exhibits amplitude variations due to spatially-varying effective loss and detuning. These parameters determine whether the quasi-CW background locally (as a function of $\theta$) exhibits the bistability that is well-known in the case of a CW pump laser [18,21], which suggests a mechanism for spontaneous single-soliton generation: as $\alpha$ is decreased, the stable effectively red-detuned branch of the resonance locally vanishes at the peak of the quasi-CW background, leading to the formation of a soliton. By following the analysis in e.g. Ref. [18] (Eqs. 11-14), we can approximate the value of $\alpha$ where this occurs at $\theta = 0$; for the diagram shown in Fig. 1b this predicts soliton generation at $\alpha = 2.741$, in excellent agreement with the value $\alpha = 2.729$ obtained in numerical simulations.

If solitons exist at $\theta \neq 0$ in a PM pumped cavity, they drift to the intracavity maximum at $\theta = 0$ [14], making superpositions of $N > 1$ solitons unstable and practically forbidden. Thus, application of PM to the pump laser removes the degeneracy between $N = 1$ and $N = 0$ and also between $N = 1$ and $N > 1$ solitons. Single-soliton generation and operation then simply requires tuning the pump power and frequency to appropriate values, regardless of initial conditions.

It is natural to consider whether a similar technique can be employed using *amplitude* modulation (AM). In principle this can be done. However, to experimentally realize the robust single-soliton operation we describe below, the AM would need to yield a narrow cavity intensity maximum, and so would necessarily be broadband (e.g. pulsed pumping [12,22]), or would require supplementary PM to provide a suitable trapping site for solitons. We are unaware of a straightforward implementation of AM for protected single-soliton operation that rivals the simplicity of the scheme presented here.

We implement the approach described above to realize completely deterministic generation of single solitons without condensation from an extended pattern, summarized in Fig. 2. We use a 22 GHz-FSR silica wedge resonator with $\Delta\nu \sim 1.5$ MHz linewidth [23], pumped by a laser with normalized power $F^2$ between 2 and 6 that is phase-modulated at a rate $f_{PM} \sim 22$ GHz with relatively small depth $\delta_{PM} \sim \pi$. This pump laser is derived from a seed CW laser using a single-sideband modulator driven by a voltage-controlled oscillator (VCO) [24]. To overcome the challenges presented by thermal instabilities [25], we also address the resonance with a counter-propagating AOM-shifted probe beam. We lock a Pound-Drever-Hall sideband of this probe beam to the resonance by feeding back to the pump-laser frequency using the VCO, which enables real-time measurement and control of the detuning $\nu_0 - \nu_{pump}$. To generate solitons we decrease the detuning from a large initial value (~40 MHz), and a soliton is generated near 5 MHz (dependent upon the pump power and coupling condition). Measuring the power converted through FWM to new frequencies, the 'comb power,' reveals a step upon soliton formation, shown in Fig. 2a. This represents a reversal of the characteristic 'soliton step' that typically signals condensation of solitons from an extended pattern and indicates direct generation of a soliton from the background. After soliton generation, $\alpha$ may be increased again without loss of the soliton, consistent with Fig. 1b. We have verified that it is possible to turn off the PM while preserving the soliton (see also Ref. [15]).

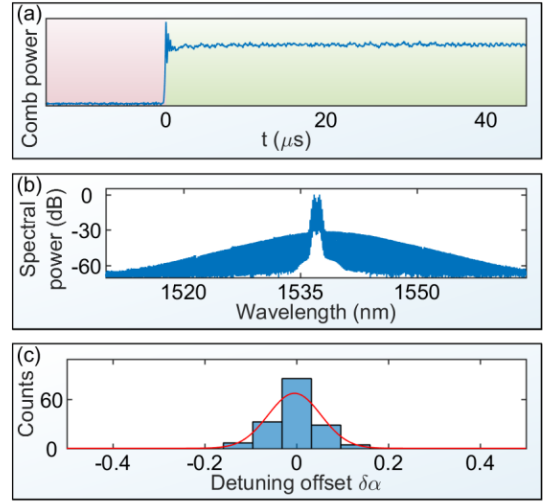

Fig. 2. (a) Measurement of a reversed soliton step in the comb power associated with soliton generation from the background. (b) Measured optical spectrum of the soliton generated with a phase-modulated pump laser. The spectrum of the pump is visible in the center. (c) Histogram of measured offset in detuning from a reference value at which a soliton is generated over 160 successive trials. The width of the interval over which solitons are generated is larger than the calculated width of the protected $N = 1$ level shown in Fig. 1b, but the model does not include laser fluctuations and other experimental effects.

Automating soliton generation by repeatedly scanning the laser into resonance (detuning ~5 MHz) and back out again (20 MHz, far enough that the soliton is lost) has enabled reversible generation of 1000 solitons in 1000 trials over 100 seconds, with a 100 % measured success rate as indicated by the comb-power trace recorded during the repeated sweep. Our probe-laser setup enables measurement of the detuning at which soliton generation occurs, which changes little from run to run. Fig. 2c presents a histogram of measurements for the generation of 160 solitons.

Besides enabling protected single-soliton operation, PM pumping also naturally provides timing and repetition-rate control, because the solitons are pushed towards the intracavity phase maximum [14]. This is illustrated in Fig. 3. In our experiments, the repetition rate of the out-coupled pulse train ($f_{rep}$) remains locked to $f_{PM}$ over a bandwidth of $\sim\pm40$ kHz. In Fig. 3a, we show a measured spectrogram of $f_{rep}$ as $f_{PM}$ is swept sinusoidally over $\pm50$ kHz. The repetition rate follows the PM except for glitches near the peaks of the sweep. In the inset of Fig. 3a we overlay the results of LLE simulations (see below) that qualitatively match the observed behavior. These simulations indicate that the periodic nature of the glitches is due to the residual pulling of the phase modulation on the soliton when the latter periodically cycles through the pump's phase maximum. Our observed locking range of $\sim\pm40$ kHz agrees well with an estimate $\delta_{PM} \times D_2/2\pi \sim 44$ kHz [14] using the approximate measured value $D_2 =14$ kHz per mode.

To investigate fast control of the repetition rate, we measure $f_{rep}$ as $f_{PM}$ is rapidly switched by $\pm40$ kHz around the soliton's natural repetition rate. We plot the resulting data as eye diagrams in Figs. 3b and 3c. In Fig. 3b, $f_{PM}$ is switched with 200 μs period and 10 μs transition time; in Fig. 3c it is switched with 100 μs period and 60 ns transition time. This data is obtained by detecting $f_{rep}$ and passing the signal through two paths, one with an element that induces a frequency-dependent phase shift. From the resulting phase difference the repetition rate can be measured in real time. These eye diagrams show that the PM enables exquisite control of the soliton pulse train.

We perform LLE simulations to further explore the dynamics of repetition-rate switching. We introduce the term $+\beta_1 \frac{\partial \psi}{\partial \theta}$ to the right-hand side of Eq. (1), where $\beta_1 = -2(FSR - f_{PM})/\Delta\nu$ represents a difference between the modulation frequency and the FSR of the resonator near the pump wavelength [14,15]; $\beta_1$ may be varied in time. In Fig. 3c we overlay a simulation of switching conducted for parameters ($\Delta\nu =1.5$ MHz, $\delta_{PM} =0.9\,\pi$) near the experimental values, and the agreement between measurements and simulation indicates that the measurements are consistent with fundamental LLE dynamics. We present the results of additional simulations in Fig. 3d; the basic observation is that the switching speed of $f_{rep}$ is limited by the resonator linewidth, and can be modestly improved by increasing $\delta_{PM}$.

One apparent barrier to the use of PM for protected single-soliton operation is the electronically inaccessible FSRs of some microcomb resonators. However, this challenge can be overcome by applying PM at a subharmonic of the FSR. Simulations indicate that solitons can be generated with small modulation depth, e.g. $\delta_{PM} = 0.15\pi$. In this limit only the first-order PM sidebands are relevant, and their amplitude and phase relative to the carrier control the dynamics. A small desired modulation depth $\delta_{PM,eff}$ defined by the relationship between the first-order sidebands and the carrier can be obtained by modulating with depth $\delta_{PM}$ at a frequency $f_{PM} \sim f_{FSR}/N$ so that the $N^{th}$-order PM sidebands and the carrier address resonator modes with relative mode numbers -1, 0, and 1, where $\delta_{PM}$ is chosen in order to achieve effective depth $\delta_{PM,eff}$. When $N$ is odd, PM is recovered when the sidebands of order $-N$, 0, and $N$ address resonator modes -1, 0, and 1. When $N$ is even pure AM results, with a driving term like $F(1 + A\cos\theta)$. Under some circumstances this AM profile also enables spontaneous soliton generation, but it cannot be obtained from a standard Mach-Zehnder modulator, which provides a drive like $F\cos(\eta + \delta\cos\theta)$.

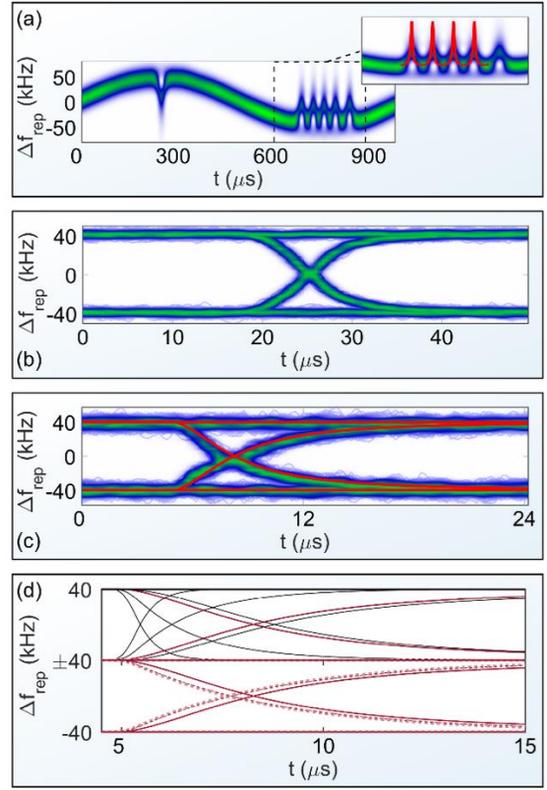

Fig. 3. (a) Measured spectrogram of $f_{rep}$ as $f_{PM}$ is swept over $\pm50$ kHz, with glitches where the locking range is exceeded. Inset: Qualitative agreement with simulations when $f_{PM}$ is outside of the locking bandwidth, shown in red. As the soliton and the pump phase evolve at different frequencies $f_{rep}$ and $f_{PM}$, the soliton periodically approaches the maximum of the phase profile. The soliton's group velocity changes, nearly locking to the phase modulation, before becoming clearly unlocked again. (b) Measured eye diagram of $f_{rep}$ as $f_{PM}$ is switched $\pm40$ kHz with 10 μs transition time. (c) The same with 60 ns transition time, and an LLE simulation of the dynamics (red) with depth $\delta_{PM} =0.9\pi$ and resonator linewidth $\Delta\nu =1.5$ MHz. (d) Simulated switching dynamics for various linewidths and modulation depths. Top, solid black: $\Delta\nu =10$ MHz (fastest, left), 3 MHz, and 1.4 MHz. Bottom, dashed red: depths of $2\pi$ and $6\pi$ (curves nearly overlap). In each, the other parameter matches the simulation in (c), shown again in solid red.

Fig. 4 presents an example of this technique. We simulate protected single-soliton generation with PM at $f_{PM} = f_{rep}/21$. The effective modulation depth is $0.15\pi$, with modulation at $f_{PM}$ of depth $\delta_{PM} \sim 8.3\pi$ ($\delta_{PM,eff} = 0.15\pi$ can be recovered with smaller $\delta_{PM}$, but the spectral efficiency is far lower). The phase modulation spreads the optical power into the PM sidebands, so this technique requires higher optical power for the same effective pumping strength; in this example the power must be increased by ~15.6 dB. While the required modulation depth and pump power are higher with subharmonic PM, neither is impractical. This technique could be used for protected single-soliton generation in high-repetition rate systems; the example above indicates that it could be immediately applied to deterministic single-soliton generation in a 630 GHz-FSR resonator with 30 GHz phase modulation.

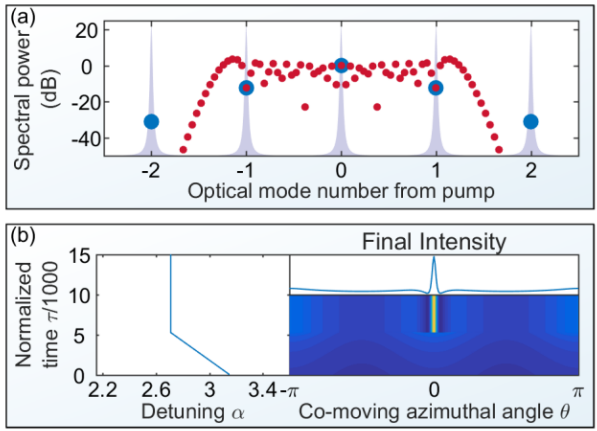

Fig. 4. (a) Spectra of PM at $f_{FSR}$ with depth $0.15\pi$ (blue) and at $f_{FSR}/21$ with depth $\sim 8.3\pi$ (red). The relationships between the fields that address resonator mode numbers -1, 0, and 1 (as indicated by the gray Lorentzian curves) are the same in both cases. (b) LLE simulation of single-soliton generation using the subharmonic phase-modulation spectrum shown in red in panel (a). Only modes $n = 0, \pm 21, \pm 42, ...$ of the phase-modulated driving field are coupled into the resonator and affect the LLE dynamics, with modes $|n| > 21$ having negligible power. As $\alpha$ is decreased from a large initial value, a soliton is spontaneously generated, exactly as in the case of phase modulation near the FSR.

In this work, we have shown that PM-pumping fundamentally changes a resonator's excitation spectrum and enables a new regime of protected single-soliton operation. The technique is applicable to resonators with electronically accessible $f_{rep}$, which are important components of proposals for photonic integration of Kerr-solitons [26,27], and can reach higher repetition-rate systems via subharmonic modulation. After soliton generation, the PM can optionally be turned off, recovering the properties of the non-PM soliton. We expect this technique to enable new experiments. For example, PM-pumped solitons are generated with known absolute timing, enabling immediate transduction of the modulation phase onto the pulse train; this is impossible with solitons stochastically condensed from an extended pattern. Our work brings microresonator solitons closer to applications.

**Funding.** NASA, DARPA DODOS, NSF DGE 1144083, AFOSR FA9550-16-1-0016, NIST, and RSNZ.

**Acknowledgement.** We thank Su-Peng Yu and Hojoong Jung for comments on the manuscript, and Andrew Weiner for helpful discussions. This work is a contribution of the US government and is not subject to copyright in the United States.